\documentclass[11pt]{aastex}

\usepackage{epsfig}
\usepackage{amsmath}
\usepackage{graphicx}
\usepackage{amssymb}
\usepackage{natbib}
\usepackage{epstopdf}
\usepackage{subfigure}
\begin{document}         
\title{Analytical Galaxy Profiles for Photometric and Lensing Analysis}
\author{David N. Spergel \\ Department of Astrophysical Sciences, Princeton University,\\ Princeton NJ 08544 USA\\
Institute of Physics and Mathematics of the Universe (IPMU),\\ Kashiwa, Japan}
\date{\today}

% Usually omit these for ApJ or MNRAS style files:
%\tableofcontents
%
%\listoffigures
%
%\listoftables

\begin{abstract}
This article introduces a family of analytical functions of the form $x^\nu K_\nu(x)$, where $K_\nu$ is the incomplete Bessel function of the 
third kind.  This family of functions can describe the density profile, projected and integrated light profiles and the gravitational 
potentials of galaxies. For the proper choice of parameters, these functions accurately approximate Sersic functions over a range
of indices and are good fits to galaxy light profiles. With an additional parameter corresponding to a galaxy core radius, these functions can  fit galaxy like M87 over a factor of $10^5$ in radius.  Unlike
Sersic profiles, these functions have simple analytical 2-dimensional and 3-dimensional Fourier transforms, so they are easily convolved with spatially varying point spread function and are well suited for photometric and lensing analysis.  We use these functions to estimate the effects of seeing on lensing measurements and show that high S/N measurements, even when the PSF is larger than the galaxy effective radius, should be able to recover accurate estimates of lensing distortions by
weighting light in the outer isophotes that are less effected by seeing.

\end{abstract}
\maketitle
%Section heading
\section{Introduction}

Weak lensing observations have the potential to provide powerful new insights into
the nature of dark energy and dark matter (see e.g., \citep{Hoekstra2008}) as well
as directly the relationship between luminous and dark matter.  

Over the next few years, astronomers can anticipate very large, high quality photometric data.  It is
essential to develop image analysis techniques that can exploit this high quality data.  The analysis 
techniques must be rapid and unbiased.  Ideally, they should be nearly optimal and use
most of the information in an astronomical image.  While there has been significant progress in the past few years, astronomers have not yet converged on an approach for image analysis
\citep{Bridle2009}.

Most image analysis takes one of two approaches: (1) fit an analytical form to the light profile such as the Sersic
profile to the galaxy distribution or (2) use an orthogonal basis function to characterize the ellipticity of an image.
While Sersic profiles have proven to be remarkably successful at fitting galaxy light
profiles outside of the central cores of galaxies  (\citet{Kormendy2009}), applying them to galaxy images requires non-linear fits to the data \citep{Ngan2009}  and
computing the effects of seeing
is computationally demanding.  The later approach (e.g., \citep{Bernstein2002})
is computationally simpler and mathematically elegant; however, Gaussians are poor approximations to galaxies.  

The goal of this paper is to define a basis function that combines the advantages of both approaches.
\S 2 introduces a series of functions, $u^\nu K_\nu(u)$ that have a simple representation in Fourier space, so can
be easily used in image analysis, and are good approximation to Sersic profiles and more importantly to galaxy photometry.  \S 3 generalizes
these functions to triaxial systems.  \S 4 applies these functions for galaxy photometry and \S 5 considers the effects of
seeing and its implications for lensing measurements.

\section{Analytical Functions for Starlight Profiles}

We consider a family of models whose 3-d Fourier profile has
the form:
\begin{equation}
\rho_\nu(k) = \frac{L_0}{4\pi^2\left[1 + (k_x^2+k_y^2+k_z^2) \left(\frac{r_0}{c_\nu}\right)^2\right]^{1+\nu}}
\end{equation}
where $r_0$ is the half-light radius, $c_\nu$ is a constant given in table 1 and $\nu > -1$.
Projecting this to two dimensions, the Fourier transform retains its  simple form:
\begin{equation}
\Sigma_\nu(k) = \frac{L_0}{2\pi\left[1 + (k_x^2+k_y^2) \left(\frac{r_0}{c_\nu}\right)^2\right]^{1+\nu}}
\end{equation}

Moving to real space, these profiles correspond to analytical 2-dimensional profiles:
\begin{eqnarray}
\Sigma_\nu(r) &=&  \int k dk J_0(k r) \frac{L_0}{\left[1 + k^2 \left(\frac{r_0}{c_\nu}\right)^2\right]^{1+\nu}} \nonumber \\
		&=&\frac{c_\nu^2 L_0}{r_0^2} f_\nu\left(\frac{c_\nu r}{r_0}\right)
\end{eqnarray}
where
\begin{equation}
f_\nu(u) = \left(\frac{u}{2}\right)^\nu \frac{K_\nu(u)}{\Gamma(\nu+1)},
\end{equation}
and $K_\nu(u)$ is a modified spherical Bessel function of the third kind.  
The Appendix describes some useful properties of these functions.

For $\nu=j+1/2$, this has a simple form:
\begin{equation}
f_{-1/2}(u) = \frac{\exp(-u)}{u}
\end{equation}
\begin{equation}
f_{1/2}(u) =  \exp(-u)
\end{equation}
\begin{equation}
f_{3/2}(u) = \frac{1}{3} \exp(-u) \left(u +1\right)
\end{equation}
At small $u$, $f_\nu(u) \to 1/(2 (\nu+1))$ for $\nu  \ge 0$ and $f_\nu(u) \to u^{2 \nu}/(2(\nu+1))$ for $-1 <\nu < 0$.  At large $u$,
$f_\nu(u) \to  \sqrt{\pi/2} \exp(-u) u^{-1/2+\nu}$.

\begin{figure}[htbp]
\begin{center}
\subfigure{\includegraphics[width=3.0in]{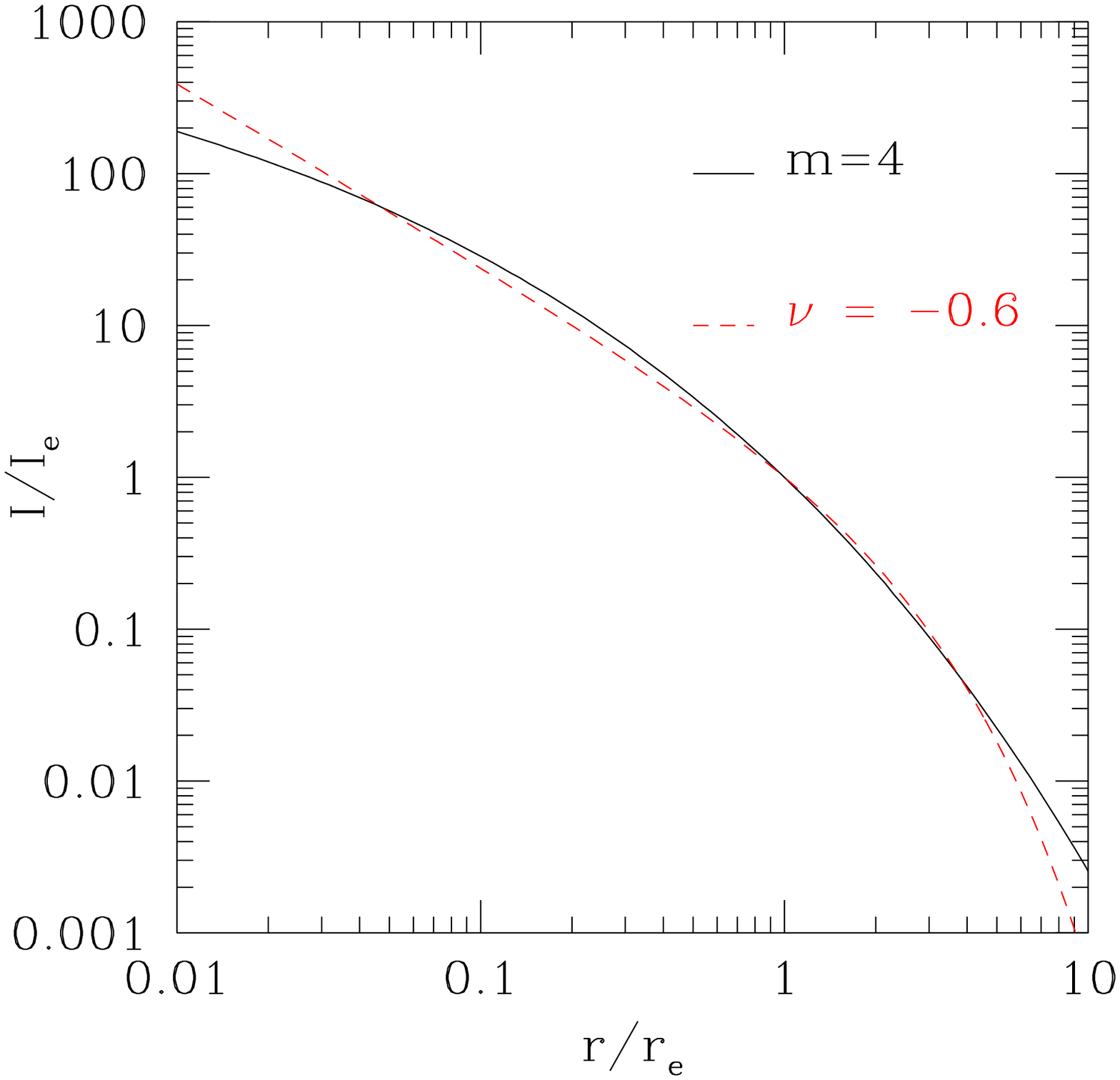}}
\subfigure{\includegraphics[width=3.0in]{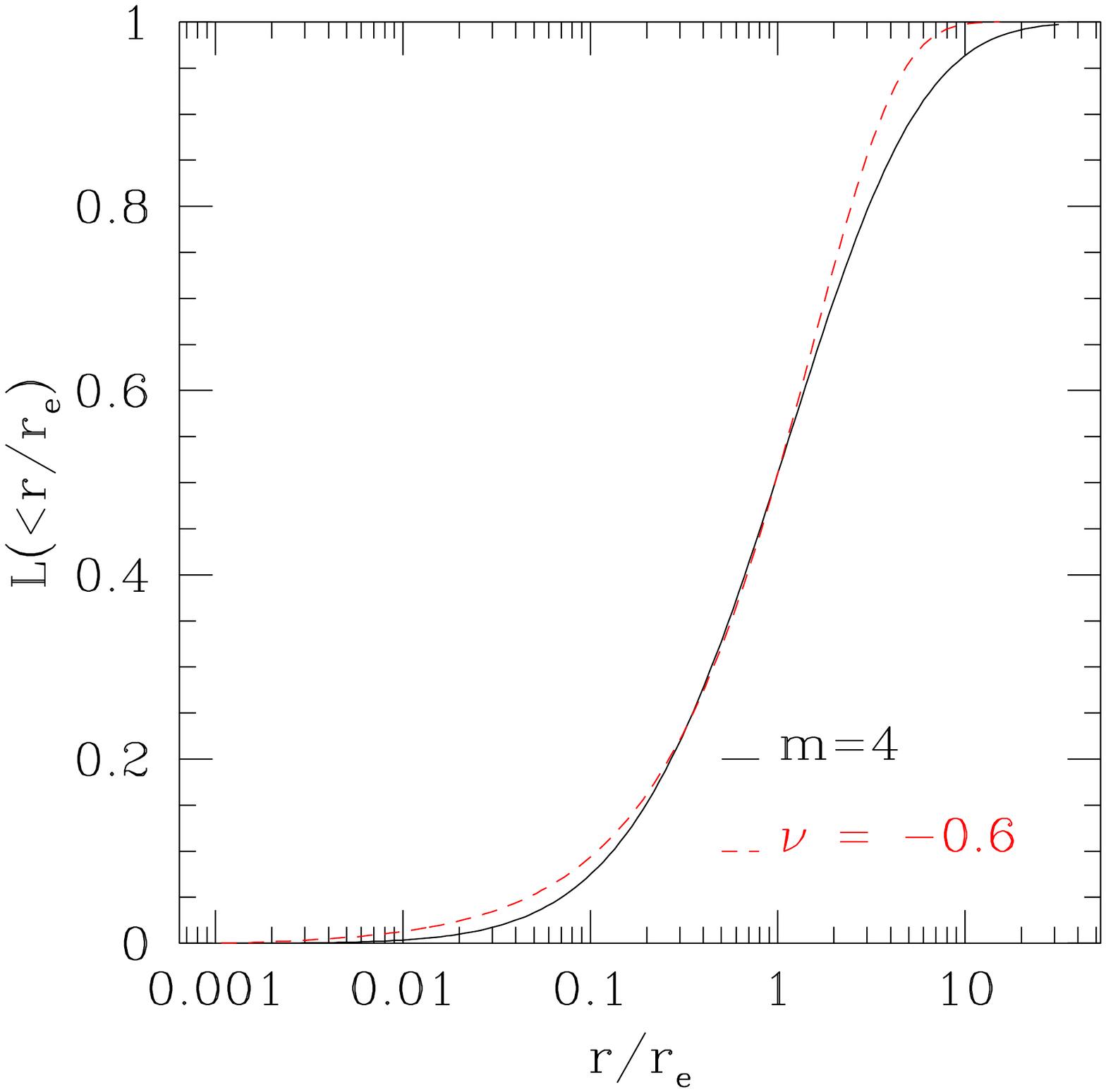}}
\vskip -0.6in
\caption{\bf This pair of figures shows that  a $\nu =-0.6$ density profile and a deVacouleur's law profile ($m=4$ Sersic profile) have
similar forms. The left panel compares the surface density profile for the two models.  The right panel compares the integrated surface
density profiles(Equation (\ref{eq:half_light})). }
\label{fig:m4}
\end{center}
\end{figure}

By going back to Fourier space and integrating by parts, we can analytically evaluate the integrated light profile of the
one component model,
\begin{eqnarray}
L(r') &=& 2\pi \int_0^{r' }\Sigma(R) RdR \nonumber \\
&=& L_0 \int_0^{r'} RdR \int_0^\infty k dk  \frac{J_0(k R)}{\left[1 + k^2 \left(\frac{r_0}{c_\nu}\right)^2\right]^{1+\nu}} \nonumber \\
&=&   L_0  r' \int_0^\infty  dk  \frac{J_1(kr')}{\left[1 + k^2 \left(\frac{r_0}{c_\nu}\right)^2\right]^{1+\nu}} \nonumber \\
&=&   L_0 \left\{1 - \frac{2(1+\nu) r_0^2}{c_\nu^2}\int_0^\infty  \frac{ kdk J_0(k r')}{\left[1 + k^2 \left(\frac{r_0}{c_\nu}\right)^2\right]^{2+\nu}} \right\} \nonumber \\
&=&  L_0\left[1 - 2 (1+\nu) f_{\nu+1} \left(\frac{c_\nu r'}{r_0}\right)\right]
\label{eq:half_light}
\end{eqnarray}
Using equation (\ref{eq:half_light}), we can solve $(1+\nu) f_{\nu+1}(c_\nu) = 1/4$ and determine the half-light
radius for the distribution.  
Table 1 lists the values of the $c_\nu$  so that  $r_0$ is the half-light radius for the light distribution.

\begin{table}[htdp]
\caption{Half-light Radii}
\begin{center}
\begin{tabular}{||c|c||c|c||c|c||c|c||}
\hline
$\nu$ & $c_\nu$ &$\nu$ & $c_\nu$ &$\nu$ & $c_\nu$ &$\nu$ & $c_\nu$ \\
\hline
-0.90 &  0.03502 &-0.45 & 0.76039 & 0.00 & 1.25715 & 0.45 & 1.64014 \\
-0.85 &  0.11212 &-0.40 & 0.82451 & 0.05 & 1.30390 & 0.50 & 1.67835 \\
-0.80 &  0.20379 &-0.35 & 0.88584 & 0.10 & 1.34943 & 0.55 & 1.71585 \\
-0.75 &  0.29616 &-0.30 & 0.94468 & 0.15 & 1.39383 & 0.60 & 1.75270 \\
-0.70 &  0.38480 &-0.25 & 1.00128 & 0.20 & 1.43717 & 0.65 & 1.78891 \\
-0.65 &  0.46864 &-0.20 & 1.05585 & 0.25 & 1.47952 & 0.70 & 1.82453 \\
-0.60 &  0.54771 &-0.15 & 1.10856 & 0.30 & 1.52093 & 0.75 & 1.85957 \\
-0.55 &  0.62240 &-0.10 & 1.15960 & 0.35 & 1.56148 & 0.80 & 1.89406 \\
-0.50 &  0.69315 &-0.05 & 1.20909 & 0.40 & 1.60120 & 0.85 & 1.92803 \\ 
\hline
\end{tabular}
\end{center}
\label{default}
\end{table}%

We can also evaluate
the 3-dimensional density profile,
\begin{eqnarray}
\rho_\nu(r) &=& \int_0^\infty k^2 dk j_0(kr) \frac{L_0}{\pi \left[1 + k^2 \left(\frac{r_0}{c_\nu}\right)^2\right]^{1+\nu}} \nonumber \\
		&=& \frac{L_0 c_\nu^3}{r_0^3}\frac{\pi^{1/2}  \Gamma\left(\nu+\frac{1}{2}\right)}{2 \Gamma(\nu+1)}  f_{\nu-\frac{1}{2}}\left(\frac{c_\nu r}{r_0}\right).
		\end{eqnarray}
		
We can evaluate the potential,
\begin{equation}
\Phi_\nu(r) = -2GM \int_0^\infty  dk \frac{j_0(kr) dk}{\pi \left[1 + k^2 \left(\frac{r_0}{c_\nu}\right)^2\right]^{1+\nu}} 
\end{equation}
by first multiplying by $r$ and differentiating,
\begin{eqnarray}
\frac{\partial}{\partial r} (r \Phi_\nu(r)) &=& -2GM \int_0^\infty  \frac{\cos(kr) dk}{ \pi \left[1 + k^2 \left(\frac{r_0}{c_\nu}\right)^2\right]^{1+\nu}}
					\nonumber \\
			&=& -GL_0 \frac{2 \Gamma(\nu+3/2)}{\sqrt{\pi} \Gamma(\nu+1)} f_{\nu+1/2}	\left(\frac{c_\nu r}{ r_0}\right)		
\end{eqnarray}
and then integrating again with respect to r to find,
\begin{eqnarray}
\Phi_\nu(r)& = - GL_0 &\left[ K_{\nu+1/2}\left(\frac{c_\nu r}{r_0}\right) L_{\nu-1/2}\left(\frac{c_\nu r}{r_0}\right) \right. \nonumber \\
		&&\qquad \left. +K_{\nu-1/2}\left(\frac{c_\nu r}{r_0}\right) L_{\nu+1/2}\left(\frac{c_\nu r}{r_0}\right) \right]
\end{eqnarray}
where $L_\nu$ is a modified Struve function.

For $\nu = 0$, this reduces to a simple form::
\begin{equation}
\Phi_0(r) = -\frac{GL_0}{r}\left[1- \exp(-c_0 r/r_0)\right]
\end{equation}
Since $\nu =0$ is intermediate between an elliptical galaxy profile ($\nu =-0.6$) and an exponential proflie ($\nu =0.5$), it is a simple potential for
a plausible generic stellar model.

We can write the potential in an alternative form by recalling the large $x$ expansion of $K_\nu(x)$:
\begin{equation}
K_\nu(x) = \sqrt{\frac{\pi}{2 x}}\exp(-x) \left[1 + \sum_j \frac{\Pi_{k=1}^j [4 \nu^2 - (2j-k)^2]}{j! (8x)^j} \right]
\end{equation}
and one of the definitions of the incomplete Gamma functions,
\begin{equation}
\int_a^\infty x^{\nu-j} \exp(-x) dx = \Gamma(\nu-j+1,a)
\end{equation}
Thus,
\begin{eqnarray}
\Phi(r)& =& -\frac{GL_0}{r} \frac{\sqrt{2}}{\Gamma(\nu+1)} \left[\Gamma\left(\nu+1,\frac{c_\nu r}{r_0}\right) \right.\\
&& \left.  +  \sum_j \frac{w_j \Gamma\left(\nu-j+1,\frac{c_\nu r}{r_0}\right)}{8^j j!}
						\right] \nonumber 
\end{eqnarray}
where $\Gamma(\alpha,x)$ is the incomplete Gamma function and $w_j = \Pi_{k=1}^j [(2\nu+1)^2 - (2j-k)^2]$.
\begin{figure}[htbp]
\begin{center}

\includegraphics[width=5in]{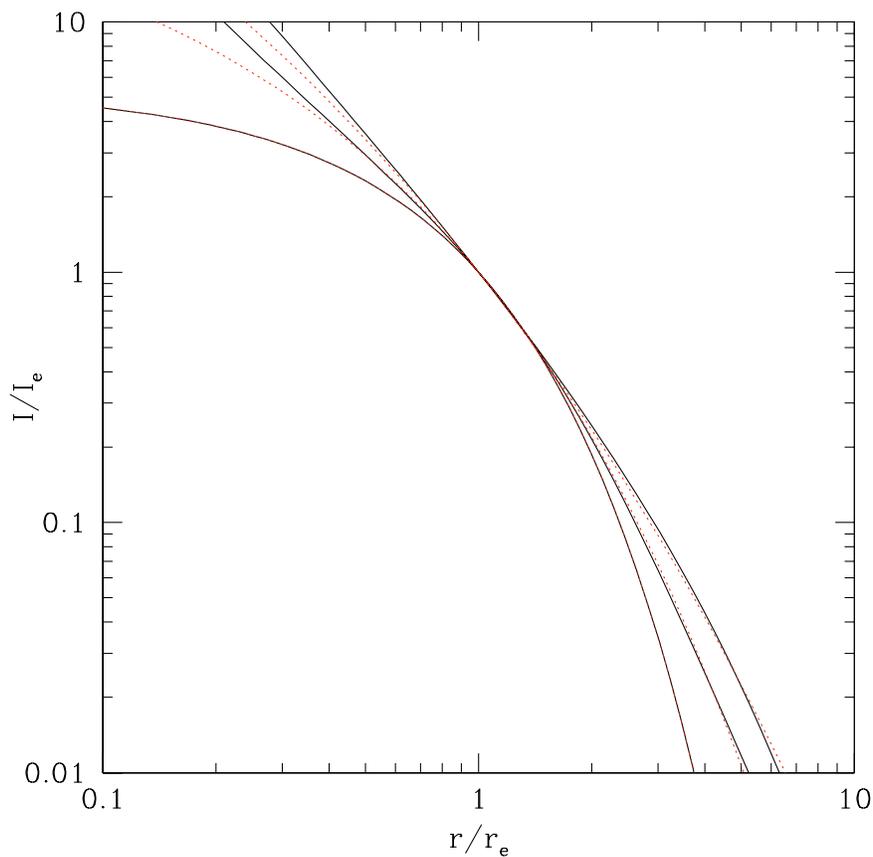}
\caption{\bf This figure compare a linear combination
of the  modified Bessel function fit (solid lines), $\beta f_{0.5}(c_{0.5} r/r_0) + (1-\beta) f_{-0.85}(0.35 r/r_0)$,
to the family of Sersic profiles.  The solid lines shows the density profiles
 for $\beta_1 = 0, 0.5$ and $1$.  The dashed lines show 
  a set of Sersic profiles  with $mn=1, 2$ and $4$.  The $n=1$ Sersic
 profile and the $\beta_1 =1$ profile are identical. }
\label{fig:Sersic}
\end{center}
\end{figure}

There are several interesting extensions to the one component representation:

(1) {\it Fixed $\nu$ models}
An alternative to using the index of the functions ($\nu$) as one of the parameters in the galaxy fit is
to represent the galaxy light profile as a sum of an exponential term $(\nu = 0.5)$ and a
very extended profile selected to fit galaxies like M87 ($\nu = 0.85$):
\begin{eqnarray}
\Sigma(r/r_0) &=&  L_1 \exp\left[-1.6783 \left(\frac{r}{r_0}-1\right)\right] \nonumber \\
&& + L_2 \left(\frac{r}{r_0}\right)^{-0.85} 
\frac{K_{0.85}(0.35 r/r_0)}{K_{0.85}(0.35)}
\label{eq:fit_Sersic}
 \end{eqnarray}
where $r_0$ is fit to the scale length of a given galaxy.    The coefficient $0.35$ was selected so that
the two profiles span the same ranges as the Sersic functions, so that most galaxy profiles
can be represented as a sum of the two functions with the same value of $r_0$. 
Figure  \ref{fig:Sersic} compares Sersic profiles with $m = 1, 2,$ and $4$ to surface
profiles fit with different linear combinations of the two terms in Equation (\ref{eq:fit_Sersic}).
While the two terms are normalized
to have the same amplitude at $r = r_0$, the later term contains more mass for $\beta =0.5$.
Note that this fitting function has three free parameters $(L_1, L_2, r_0)$, the same number of parameters as the Sersic
profile $(L ,n, r_0)$ and the profile introduced earlier in this section $(L,\nu, r_0)$.  The main advantage of
Equation (\ref{eq:fit_Sersic}) is apparent in \S \ref{sec:fitting} where we show that we can precompute the effects of atmospheric seeing.

\begin{figure}[htbp]
\begin{center}
\vskip -0.85in
\includegraphics[width=4.5in]{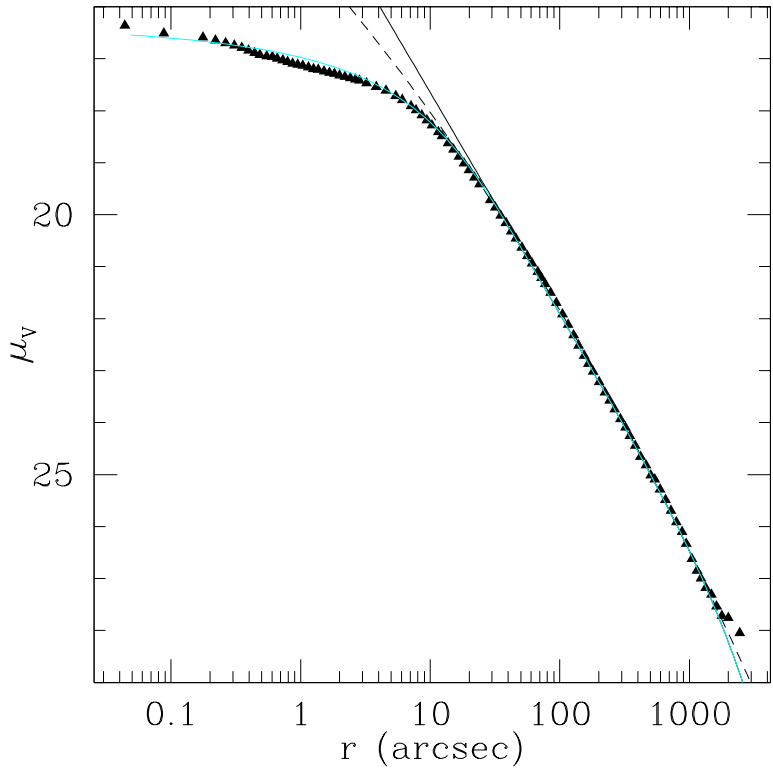}
\vskip -1.0in
\caption{\bf This figure compares Kormendy et al. (2009) compilation of observations of M87 (NGC 4468)
to their best fit Sersic profile (dashed line with $m=11.8$ and $r_e = 703.91 {\rm kpc}$), a one component profile (solid line with $\nu = -0.85$ and
 $r_0/c_\nu= 1944 {\rm kpc}$) and a two component profile (cyan solid line with $r_0/c_\nu= 1944 {\rm kpc}$, $\nu=-0.85$, and $r_{core}/r_0 =0.0043$).}
\label{fig:M87}
\end{center}
\end{figure}

(2) {\it Core Radii}
For galaxies with well defined cores  but extended light profiles (see Figure \ref{fig:M87}), we can generalize the $\nu < 0$ profile by adding a second component
and an additional parameter, $r_{core}$:
\begin{equation}
\Sigma(r) = \frac{L_0 c_\nu^2}{r_0^2\left[1 -  \left(\frac{r_{core}}{r_0}\right)^{2(1-\nu)}\right]} \left[f_\nu\left(\frac{c_\nu r}{r_0}\right) - \left(\frac{r_{core}}{r_0}\right)^{2\nu}
 f_\nu\left(\frac{c_\nu r}{r_{core}}\right) \right]
\label{eq:two_comp}
\end{equation}
This two component form approaches an exponential for $r < r_{core} < r_0$.  This generalized surface brightness distribution also has a simple
representation in Fourier space:
\begin{equation}
\Sigma(k) = \frac{L_0}{r_0^{2\nu}-r_{core}^{2\nu}}\left[\frac{r_0^{2\nu}}{\left(1+k^2 r_0^2/c_\nu^2\right)^{1+\nu}} -  \frac{r_{core}^{2\nu}}{\left(1 + k^2 r_{core}^2/c_\nu^2\right)^{1+\nu}} \right]
\end{equation}

(3) {\it Triaxial Galaxies}.
These profiles can be generalized to triaxial galaxies by introducing a a change of
variables to a new set of coordinates, $\vec r= {\bf F} \vec u$ and $\vec q = {\bf F}^{-1} \vec s$,
so that we can rewrite our profiles,
\begin{equation}
g(\vec r) = \int d^nq\  \tilde g(\vec q) \exp(i \vec q \cdot \vec r) ,
\end{equation}
in a new set of coordinates:
\begin{equation}
g({\bf F}\vec u) = \det|{\bf F}^{-1}| \int d^ns\  \tilde g({\bf F}^{-1} \vec s) \exp(i \vec u \cdot \vec s) 
\label{eq:cov}
\end{equation}

Applying the transformation, $k_x \to k_x \sqrt{1-\epsilon}$ and $k_y \to k_y\sqrt(1+\epsilon)$  the 
 two dimensional light distribution in Fourier space becomes,
\begin{equation}
\Sigma_\nu(\vec k) = \frac{L_0}{\left[1 + k^2 \left(\frac{r_0}{c_\nu}\right)^2\ (1- \epsilon \cos(2\phi_k))\right]^{1+\nu}} 
\end{equation}
which has a corresponding real space representation:
\begin{equation}
\Sigma_\nu(\vec R)= \frac{L_0 c_\nu^2}{r_0^2} f_\nu\left[\frac{c_\nu r}{r_0 \sqrt{1 -\epsilon^2}}
\sqrt{1 + \epsilon \cos(2\phi_r)}
			\right]
\end{equation}

We can expand out the Fourier space profile in a Taylor Series:
\begin{equation}
\Sigma_\nu(\vec k) = L_0\sum_j \frac{\Gamma(2+\nu)}{\Gamma(2+\nu-j)j!} \frac{\left[\epsilon k^2 \left(\frac{r_0}{c_\nu}\right)^2\cos(2\phi_k))\right]^{j}}{\left(1+k^2 \left(\frac{r_0}{c_\nu}\right)^2\right)^{1+\nu+j}}
\label{eq:taylor}
\end{equation}

\section{The Effect of Seeing and Finite Resolution}

In this section, we use the profiles to explore the effects of atmospheric seeing  and
telescope distortions on determinations of ellipticity.  

The observed image, $I(\vec r)$,  is a convolution of the galaxy's surface brightness
profile, $\Sigma(\vec r)$ with the effects of the atmosphere
and the telescope optics:,
\begin{equation}
I(\vec r) = \int d^2k \exp(i \vec k \cdot \vec r) \Sigma(\vec k) R_{turb}(\vec k) R_{telescope}(\vec k)
\end{equation}
where $R_{turb}$ is the effect of atmospheric turbulence and $R_{telescope}$ is the
response function of the telescope.
Using Kolmogorov turbulence theory,
\begin{equation}
R(k) = \exp(-(kb)^{5/3})
\end{equation}
where $\theta_{FWHM} = 2.9207 b$ and $b$ is a parameter that characterizes the correlation length of the 
atmospheric turbulence.  Since most astronomical observations are seeing-limited,
we set $R_{telescope}=1$ for this section.

Seeing reduces the signal/noise of the fits of galaxy (and star) profiles to the observations.
We can estimate the signal to noise by considering fitting a multi-parameter model to 
the data:
\begin{equation}
\chi^2 = \sum_i  \frac{\Delta \Omega}{n_0^2} [I_{measured}(r_i) - I_{model}(\vec A, r_i)]^2
\end{equation}
where $n_0$ is the noise times unit area, $\Delta \Omega$ is the
area of the pixel, $r_i$ is the pixel centroid, and $I_{measured}$ is the measured light profile and
$I_{model}$ is a model with a number of parameters $\vec A$.  The noise estimate assumes that we are observing faint galaxies and are dominated by atmospheric noise (and/or read noise).

The simplest case is a point source of intensity $A$:
\begin{equation}
\chi^2 = \sum_i \frac{\Delta \Omega}{n_0^2} [I_{measured}(r_i) - A I_{PS}(r_i)]^2
\end{equation}
.  The signal-to-noise for point source detection is:
\begin{equation}
\left(\frac{S}{N}\right)^2_{PS} =\frac{\partial \chi^2}{\partial A^2} = \sum_i \frac{I_{PS}^2(r_i) \Delta \Omega}{n_0^2}
\end{equation}
In the continuum limit, we can use Parseval's theorem to evaluate the signal-to-noise:
\begin{eqnarray}
\left(\frac{S}{N}\right)^2_{PS} &=& \frac{1}{n_0^2} \int d^2r I^2_{PS}(r) \nonumber \\
			&=&  \frac{1}{n_0^2} \int d^2k R^2_{a}(\vec k) \equiv \frac{P_{PS}}{n_0^2}
\end{eqnarray}

\begin{figure}[htbp]
\begin{center}
\subfigure{\includegraphics[width=3.2in]{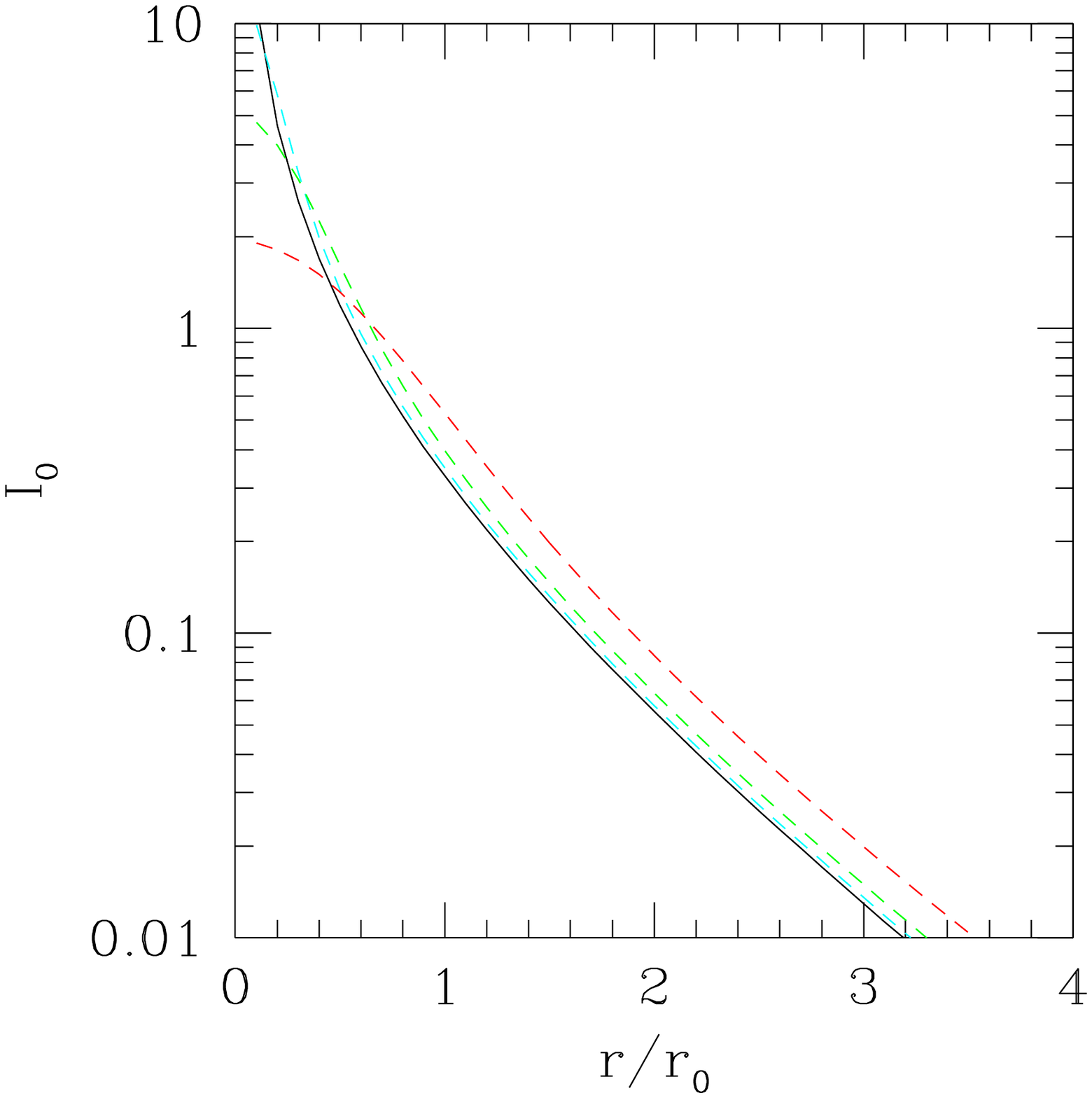}}
\subfigure{\includegraphics[width=3.2in]{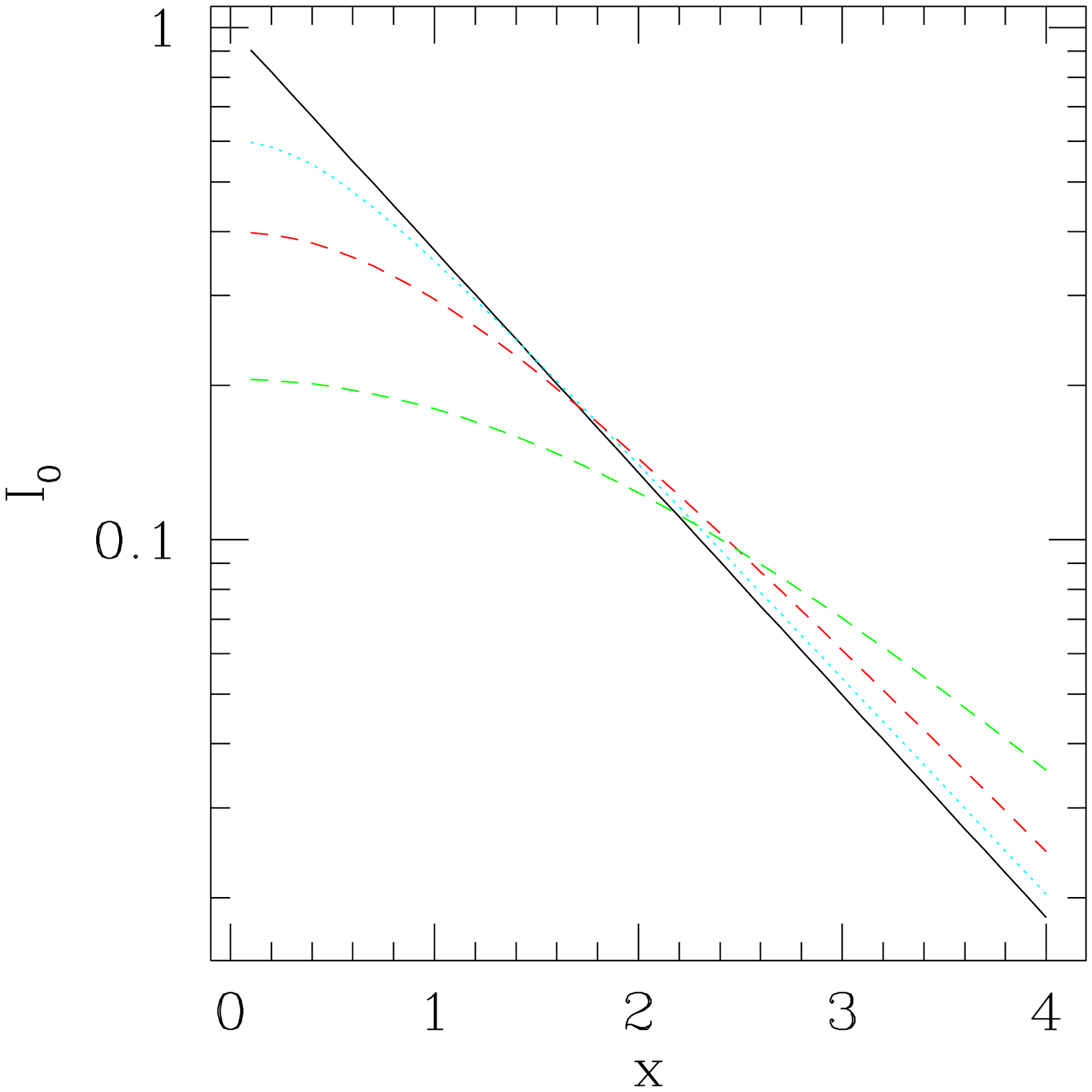}}
\vskip -0.6in
\caption{\bf This shows the effects of seeing on the galaxy profile.  The four lines in the left panel show the amplitude of the
monopole term (Eq. \ref{eq:mono}) for $\theta_{FWHM}/r_0 = 0, 0.5, 1.0$ and $2.0$.
The left panel is  for an elliptical
profile ($\nu =-0.6$) and the right panel is for an exponential profile ($\nu = 0.5$).}
\label{fig:exp}
\end{center}
\end{figure}

\begin{figure}[htbp]

\begin{center}
\subfigure{\includegraphics[width=3.2in]{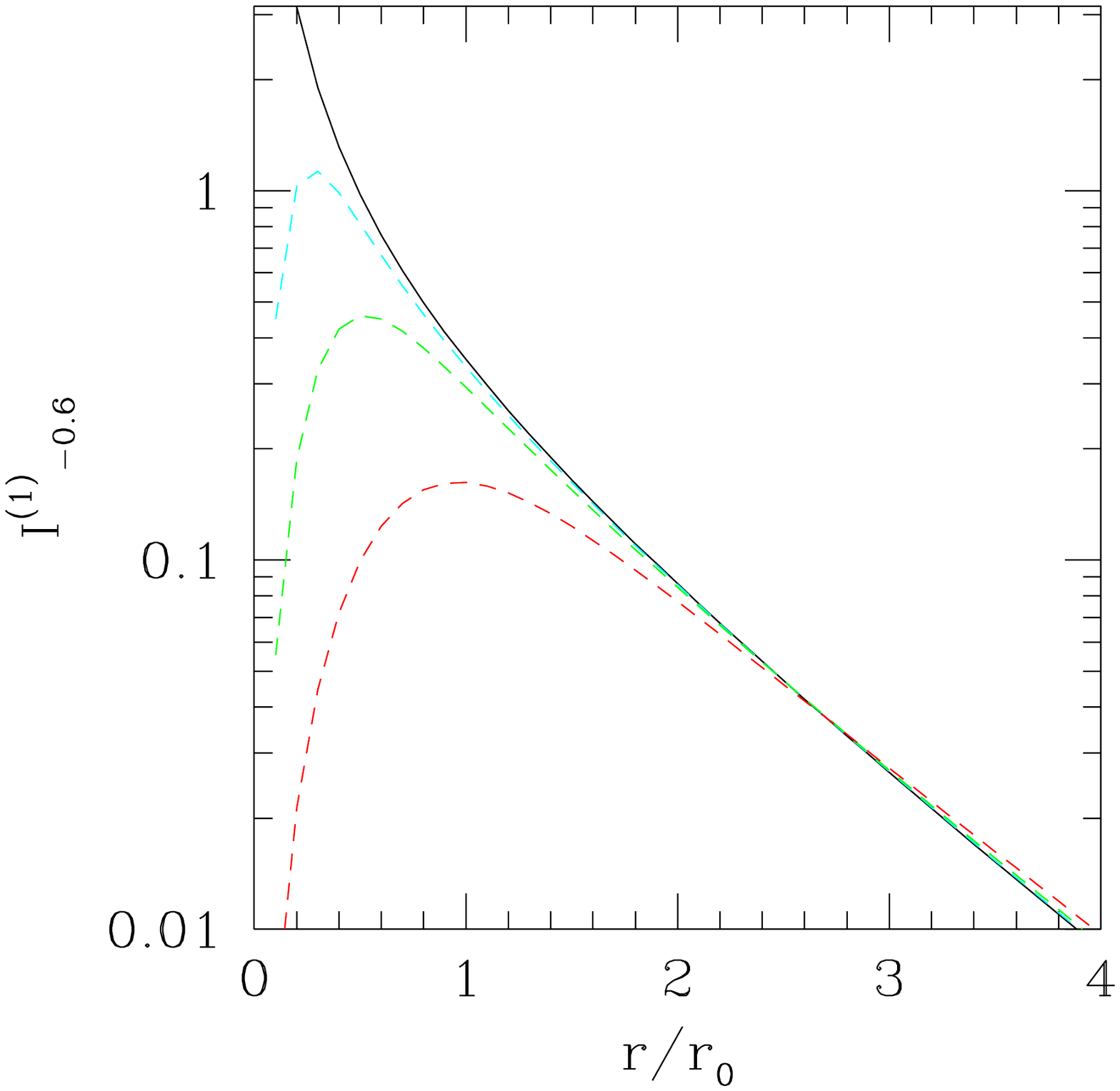}}
\subfigure{\includegraphics[width=3.2in]{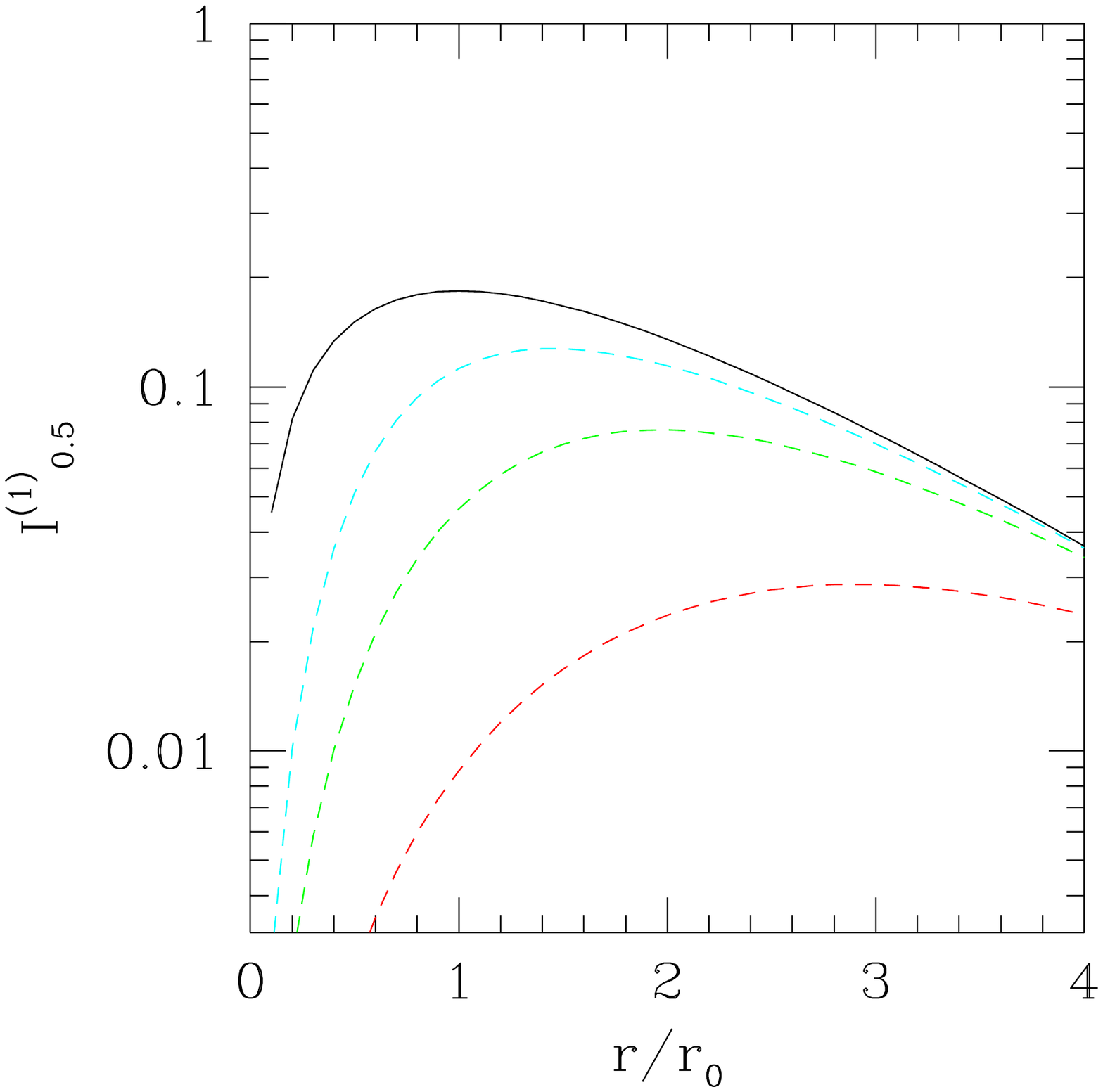}}
\vskip -0.6in
\caption{\bf This shows the effects of seeing on the galaxy profile.  The four lines in the left panel show the amplitude of the
quadrupole  term (Eq. \ref{eq:dipole}) for $\theta_{FWHM}/r_0 = 0, 0.5, 1.0$ and $2.0$.
The left panel is  for an elliptical
profile ($\nu =-0.6$) and the right panel is for an exponential profile ($\nu = 0.5$)}
\label{fig:ell}
\end{center}
\end{figure}

For a circularly symmetric galaxy fit with the profile of section 2, the observed profile can again be calculated
by the convolution integral:
\begin{eqnarray}
I_\nu^{(0)}(r) &=& \int \frac{d^2k \exp\left[-i\vec k\cdot \vec r\right] \exp(-(kb)^{5/3})}
						{\left(1+k^2 \left(\frac{r_0}{c_\nu}\right)^2\right)^{1+\nu}}
 \nonumber \\
				&=& \frac{ c_\nu^2}{r_0^2}
				\int \frac{ \tilde k d \tilde k J_0(\tilde kr) \exp\left[- \left( \frac{\tilde k b
				c_\nu}{r_0}\right)^{5/3}\right]}{\left(1+\tilde k^2\right)^{1+\nu}}
				\label{eq:mono}
\end{eqnarray}
where $\tilde k = k r_0/c_\nu$.
The effect of atmospheric seeing is a function of $b/r_0$ or equivalently a function of the ratio of the FWHM of the PSF to
the effective radius of the galaxy.  Figures \ref{fig:exp} and \ref{fig:ell} show that atmospheric seeing``moves" light from the central cusp outwards and circularizes the inner portions of the galaxy.  These effects are less
dramatic for the outer isophotes.

The signal-to-noise for the galaxy detection is the second derivative of
$\chi^2$ with respect to $M$:
\begin{equation}
\left(\frac{S}{N}\right)^2_{M}  = \left(\frac{S}{N}\right)^2_{PS} \frac{P_0}{P_{PS}}
\label{eq:SN1}
\end{equation}
where $(S/N)_{PS}$ is the signal-to-noise for the detection of a point source of the same magnitude.
\begin{equation}
P_0 = 
{\int  \frac{d^2\tilde k}{\left(1+ \tilde k^2\right)^{2+2\nu }}\exp\left[- 2\left( \frac{\tilde k b c_\nu}{r_0}\right)^{5/3}\right]}
\end{equation}
Note that the effect of galaxy finite size  is to degrade the S/N by the ratio of the effective areas of a seeing convolved point source to the seeing convolved galaxy.

Seeing has an even more dramatic effect on the ellipticity of the image.  The convolution of the $\cos(2\phi)$ term in the 
light profile,
\begin{eqnarray}
I_\nu^{(1)}(r) &\equiv& (1+\nu) \left(\frac{r_0}{c_\nu}\right)^2  \int \frac{d^2k \exp\left[i \vec k\cdot \vec r \right] \exp(-(kb)^{5/3}) k^2 \cos(2 \phi_k)}
						{\left(1+k^2 \left(\frac{r_0}{c_\nu}\right)^2 \right)^{2+\nu}}
 \nonumber \\
				&=& \frac{\epsilon L_0 (1+\nu) c_\nu^2}{r_0^2}   \cos(2\phi_r)  \int \frac{ \tilde k^3 d\tilde k J_2(\tilde kr) \exp\left[- \left( \frac{\tilde k b
				c_\nu}{r_0}\right)^{5/3}\right]} {\left(1+\tilde k^2\right)^{2+\nu}}
				\label{eq:dipole}
\end{eqnarray}
is shown in Figure \ref{fig:ell}.  Seeing makes the central region of the galaxy round and has a reduced effect in the outer profile.
Because the outer profile is less effected by seeing, it is useful to use a functional form that is a good fit to the outer profile for
measurements of ellipticity.  Galaxies are not well-described by Gaussians and the Hermite profile based determination of ellipticities are
"missing" the information in the outer profiles.

\begin{figure}[htbp]
\begin{center}
\includegraphics[width=4in]{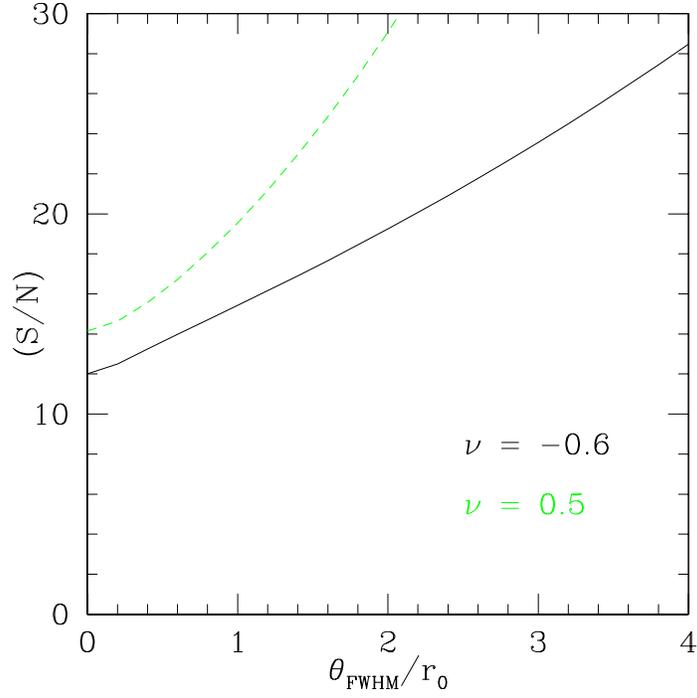}
\vskip -0.6in

\caption{\bf As the seeing degrades, longer integrations are needed to achieve the required sensitivity to shear.  This figure quantifies this effect
by plotting the point source S/N need to achieve a statistical error on the shearr
of $\sigma_{\epsilon)}= 0.2$.  This point source S/N is shown  as a function of the ratio of the FWHM of the PSF to the effective radius of the galaxy.  The solid line is 
for a $\nu = -0.6$ profile, which corresponds to an elliptical galaxy.  
The dashed line is for a $\nu = 0.5$ profile, which corresponds to an
 exponential profile. }
\label{fig:sn}
\end{center}
\end{figure}

We can again estimate the signal to noise by differentiating the fit:
\begin{equation}
\chi^2 = \sum_i \frac{1}{\sigma_i^2} [I(r_i) - L_0 I^0_\nu(r_i) - \epsilon \L_0 I^1_\nu(r_i)]^2
\end{equation}
Because the $\cos(2\phi)$ term is orthogonal to the symmetric term, the error on the
second term is again just the second derivative with respect to its amplitude:
\begin{equation}
\sigma_{\epsilon L}^{-2} =\frac{1}{2} \left(\frac{S}{N}\right)_{PS}^2 
\frac{\int r dr\left [I_\nu^{(1)}(r)\right]^2}{\int r dr\left [I_\nu^{(0)}(r)\right]^2}
\frac{ \int rdr I_\nu^{(0)}(r)} 
{\int rdr I_nu^{(1)} (r) }
\end{equation}
where the $(1/2)$ factor comes from angle averaging $\cos(2 \phi_r)$.
Using Parseval's theorem and Equations (\ref{eq:mono}- \ref{eq:dipole}):
\begin{equation}
\sigma_{\epsilon L}^{-2} = \left(\frac{S}{N}\right)_{PS}^2  \frac{P_2}{P_0}
\label{eq;SN2}
\end{equation}
where,
\begin{equation}
P_2 = \frac{1+\nu}{2} \left(\frac{c_\nu}{r_0}\right)^6
\int \frac{\tilde k^5 d\tilde k}{\left(1+ \tilde k^2\right)^{4+2\nu } }\exp\left[- 2\left( \frac{\tilde k b c_\nu}{r_0}\right)^{5/3}\right]
\end{equation}
This yields:
\begin{equation}
\sigma_{\epsilon}^{-2}  = \left(\frac{S}{N}\right)_{PS}^2  \frac{P_2 P_0}{(P_2 + P_0)P_{PS}}
\end{equation}
Figure (\ref{fig:sn}) quantifies the additional integration time needed to measure the ellipticity with an
uncertainty, $\sigma_\epsilon =1/5$ by plotting the point source signal to noise needed as a function of
the ratio of the PSF FWHM to the effective radius of the galaxy,
\begin{equation}
\left(\frac{S}{N}\right)_{PS} = 5 \sqrt{ \frac{(P_2 + P_0)P_{PS}}{P_2 P_0}}
\end{equation}

Figure (\ref{fig:exp} - \ref{fig:sn}) show the effects of atmospheric seeing on lensing measurements.  The left panels
in Figures (\ref{fig:exp}) and (\ref{fig:ell}) show that as the seeing degrades,
the image gets broader.  While the right panel shows that as the seeing degrades, the images appear rounder, particularly
in the inner regions.  Note that seeing has less of an effect on the outer isophotes.  Thus, high S/N observations can recover
accurate ellipticity measurements, even if $\theta_{FWHM} > r_0$.  However, since the information about the ellipticity is only
in the outer isophotes, it is important to use an optimized weighting scheme.

\section{Fitting the Profile to Observations\label{sec:fitting}}

This section presents an algorithm for fitting light profiles to multi-image stacked data.
The algorithm utilizes a linearized version of the two component light profile
introduced in the previous section (see equation (\ref{eq_ell_k_profile})).  Since the goal
of this approach is to have a fast linear algorithm that can be applied to large data sets, the profile
is expanded in a power series in the image size and shape and the convolution of the seeing
with each of the terms in the series is precomputed.  Since we are also expanding the PSF in
a series of terms, we can simultaneous fit for multi-frame images with little increase in computational cost.
The final step in the fit is a non-linear step that is not very computationally intensive and 
returns the intensity, size, shape, profile, and orientation of each galaxy image.

Following the approach outlined in section 2, the galaxy profile is parameterized as a five parameter fit:
the amplitude of the exponential profile, $L_1$, the ampltude of the elliptical M87-like profile term, $L_2$,
a flattening term, $\epsilon$, a scale radius $r_0$ and an orientation, $\phi_0$:
\begin{eqnarray}
\Sigma(\vec r|L_1,L_2,r_0,\epsilon,\phi_0)&=& \Sigma_1(\vec r|L_1,r_0,\epsilon,\phi_0) + \Sigma_2 (\vec r|L_2,r_0,\epsilon,\phi_0) \nonumber \\
&=&L_1 \alpha_1 f_{1/2}\left(c_{1/2} \frac{r\sqrt{1 + \epsilon \cos(2(\phi-\phi_0)}}{r_0}\right)
 \\
&& +L_2\alpha_2 f_{-0.85} \left(0.35 \frac{r\sqrt{1 + \epsilon \cos(2(\phi-\phi_0)}}{r_0}\right)\nonumber
\label{eq_ell_profile}
\end{eqnarray}
where $\alpha_1 = \exp(1.6783)=1/f_{1/2}(c_{1/2})$ and $\alpha_2 = 1/f_{-0.85}(0.35)$.
This fit has a simple representation in Fourier space:
\begin{equation}
\label{eq_ell_k_profile}
 \sum_{i=1,2} 
\frac{L_i}{\left\{1+ k^2 \gamma_i^2 r_0^2 \left[1 - \epsilon \cos(2(\phi-\phi_0)\right]\right\}^{1+\nu_i}}
\end{equation}
where  $\nu_1 = 0.5$, $\nu_2 = -0.85$,
$\gamma_1 = 1/c_{1/2}$, and $\gamma_2 = 1/0.35$.  Note that we are fitting a single value of $r_0$, the effective radius to the prorfie.

The first step in the analysis is to fit for the position of the galaxy, $\vec \theta_g$ and estimate its size from
measuring its half light radius.   By first computing the convolution of
a series of circularly symmetric galaxy profiles with different characteristic size (e.g., in steps of 0.1"), the profile fitting calculation becomes
a linear problem.  Instead of fitting for $r_0$, the effective radius, we can fit for 
\begin{equation}
\Delta \equiv  1-\left(\frac{r_0}{r_i}\right)^2
\end{equation}
where $r_i$ is one of the nearest precomputed  fit to the initial list of values .  Expanding the profile as a 
 Taylor series in ($\Delta
+(1 -\Delta)\epsilon \cos(2(\phi-\phi_0))$:
\begin{eqnarray}
\Sigma(\vec k|L_1,L_2,\Delta,,\epsilon,\phi_0) &=& \sum_i
\frac{L_i}{\left[1 + k^2 \gamma_i^2 r_0^2 (1- \Delta)(1 -\epsilon \cos(2(\phi-\phi_0)) \right]^{1+\nu_i}} \nonumber \\
&=& \sum_i \sum_{j=0} \left[\Delta+(1- \Delta)\epsilon \cos(2(\phi-\phi_0))\right]^j 
\frac{w_j(\nu_i)(\gamma_i kr_0)^{2j}}
	{\left[1+\gamma_i^2 k^2 r_0^2\right]^{\nu_i+j+1}}
\end{eqnarray}
where $w_j(\nu_i) = \Pi_{s=1}^j (\nu_i+s)/j!$.    
We then rewrite the profile as a sum of terms with linear fit coefficents:
\begin{equation}
\Sigma_i(\vec k|a_{q}) = \sum_{jm} a_{ijm}(L_1,L_2,\Delta,\epsilon,\phi_0) \mu_{ijm}(\vec k,r_i) =\sum_q a_q \mu_q(\vec k,r_i)
\end{equation}
where the sum over $i,j$ and $m$ is represented as a sum over $q$ for notational simplicity:
\begin{equation}
a_{ijm} = L_i \Delta^{j-m} (1-\Delta)^m  \epsilon^m \exp(-2 im \phi_0),
\label{eq:a}
\end{equation}
and 
\begin{equation}
\mu_{ijm}(\vec k, r_i) =   \frac{j!}{(j-m)! m!} \frac{(\gamma_i kr_i)^{2j}w_j(\nu_i) \exp(2i m \phi)}
	{\left[1+\gamma_i^2 k^2 r_i^2\right]^{\nu_i+j+1} }.
\end{equation}

The next step is to represent the time-varying and spatially varying 
 PSF can be expanded as a sum of specified (but not necessarily orthogonal functions), $P_s$
(see e.g., \citep{Jarvis2004}):
 \begin{equation}
R_{PSF}(\vec k,\vec\theta_g,t_n) = \sum_s U_s(\vec \theta_g,t_n) P_s(\vec k),
\label{eq:PSF}
\end{equation}]
where $t_n$ is the time of each frame, $\theta_g$ is the galaxy position, and
$P_s(\vec k)$ is the Fourier transform of the basis function.  Typically, $U_s(\vec \theta,t_p)$ will be fit to stars in
the image.

The image profile in a given frame can now be expressed as a sum,:
\begin{equation}
\tilde I(\vec r,t_n|\vec a_q) = \sum_q a_q \sum_s U_s(\vec \theta_g,t_n) \tilde \mu_{qs}(\vec r,r_i),
		 \label{eq:prof}
\end{equation}
over a series of terms that can be precomputed once for a series of effective radii, $r_i$:
\begin{equation}
\tilde \mu_{qs}(\vec r) = \int d^2k \exp(-i \vec k \cdot \vec r) P_s(\vec k) \mu_{q}(\vec k, r_i)
\end{equation}
These function, $\tilde \mu_{ijms}$, describe the convolution of a term in the expansion of the galaxy light profile
with a term in the expansion of the PSF.

Fitting the convolved linear profile to the observations is now a linear process
that fits the galaxy light profile coefficients,  $a_k$, to all of the data in the stack:
\begin{equation}
\chi^2 = \sum_{l,n} \frac{(I_{\rm measured}(\vec r_l,t_n) - \sum_q a_q \sum_s U_s\vec r_g,t_n) \tilde \mu_{qs}(\vec r,r_i)))^2}{\sigma_{ln}^2}
\label{eq:chisq}
\end{equation}
where the sum over $l$ is a sum over pixels.  The best fit is the solution to a linear equation,
\begin{equation}
N_{qq'}^{-1} a^{ML}_{q'} = b_{q}
\end{equation}
where 
\begin{equation}
N_{qq'} =  \sum_{ln} \frac{\tilde \mu_q(\vec r_l,t_n) \tilde \mu_{q'}(\vec r_l,t_n)}{\sigma_{ln}^2}
\end{equation}
\begin{equation}
b_q = \sum_{ln} \frac{I_{measured}(\vec r_l,t_n)\tilde \mu_q(|vec r_l,t_n)}{\sigma_{ln}^2}
\end{equation}
are evaluated as a sum over pixels and frames using precomputed functions.
This linear fit takes $N_{stack} N_{pix} N_{gal terms}^2$ steps to evaluate $N_{kk}$ and $N_{gal terms}^3$ steps to invert the matrix.
If we expand to first order in the ellipticity, then $N_{gal terms} = 6$.  If we work to fourth order, then $N_{gal terms} = 15$.
This very rapid process should enable fits to individual observations in the stack.

We then do a nonlinear fit to solve for the amplitude of the terms in the expansion, $L_1$, $L_2$, 
 $\epsilon$, $\phi_0$, and $\Delta$:
\begin{eqnarray}
\label{eq:chisq}
\chi^2(a_q|L_1,L_2,\Delta,\epsilon,\phi_0) =& \sum_{qq'}& [a^{ML}_q- a_q^{fit}(L_1,L_2,\Delta,\epsilon,\phi_0)]N_{qq'}^{-1} \nonumber \\
&&[a_{q '}^{ML}- a_{q'}^{fit}(L_1,L_2,\Delta,\epsilon,\phi_0)]
\end{eqnarray}
where $a_{q}$ is defined in Equation (\ref{eq:a}).
This operation is very quick as it takes only $5 N_{gal terms}$ steps to evaluate the non-linear fit.

\section{Conclusions and Next Steps}

This note introduces a basis function that may prove useful for the analysis of galaxy images,
particularly for lensing work.  In a subsequent paper, we will test this approach against the GREAT08 simulations \citep{Bridle2009}.

There are several possible extensions to this parameter fit.  We can include priors on $\epsilon$
and modified the $\chi^2$ (Equation (\ref{eq:chisq}) to be a likelihood function and
then marginalize over the galaxy size and position.  We could then also include
priors on the ellipticity functions\citep{Kitching2008}.  Other possible generalization would be 
to include PSF uncertainties and  to  generalize the expansion to include shapelet terms.
\section{Acknowledgements}

This research was partially supported by NSF grant AST/0908292 through the ARAA.  I would like to thank Alan Heavens, Chiaki Hikage,
Robert Lupton, Rachel Mandelbaum, Michael Strauss and Masahiro Takada for helpful comments.

\bibliography{profile_fit_apj}

\appendix

\section{Appendix: Useful Properties of Bessel Functions and  Incomplete Bessel Functions of the Third Kind}

\citet{Kostroun1980} provides a useful expression for the numerical evaluation of $K_\nu(u)$:
\begin{equation}
K_\nu(u) = h\left[ \frac{\exp(-u)}{2} + \sum_{r=1}^\infty \exp(-u \cosh(rh))\cosh(\nu r h) \right]
\end{equation}
where for values of $\nu$ of interest to this paper, we set $h=0.5$ and find that the sum
converges to $10^{-10}$ with less than 7 terms for most values of $\nu$ and $u$ used in the paper.

There are useful relationships between various modified Bessel functions:
\begin{equation}
\int J_m(kr) \frac{k^{m+1 }dk}{\left(1+k^2\right)^{1+m+\nu}} =\left(\frac{r}{2}\right)^{m+\nu}\frac{ K_\nu(r)}{\Gamma(1 + m + \nu)}
\end{equation}

\begin{eqnarray}
\int j_m(kr) \frac{k^{m+2} dk}{\left(1+k^2\right)^{1+m+\nu}} &=&\left(\frac{r}{2}\right)^{m+\nu-1/2}\frac{\sqrt{\pi} K_{\nu-1/2}(r)}{\Gamma(1 + m + \nu)}
								\nonumber \\
										&=& \left(\frac{r}{2}\right)^{m} f_{\nu-1/2}(r) \frac{\sqrt{\pi} \Gamma(\nu+3/2)}{\Gamma(1+m+\nu)}
\end{eqnarray}
For $\nu < 0$, we can use \citet{AS1972} equation 9.6.25,
\begin{eqnarray}
f_\nu(z) &=& \frac{\sqrt{\pi}}{\Gamma(\nu+1)\Gamma(\nu+1/2)}\int_0^\infty \exp(-z\cosh t) \sinh^{2\nu} t dt \nonumber \\
	     &=&  \frac{\sqrt{\pi}}{\Gamma(\nu+1)\Gamma(\nu+1/2)}\int_1^\infty \frac{\exp(-zu) }{\left(u^2 - 1\right)^{1/2-\nu}} du
\end{eqnarray}
Thus,
\begin{equation}
\int_x^\infty f_\nu(z) dz =  \frac{\sqrt{\pi}\exp(-x}{\Gamma(\nu+1)\Gamma(\nu+1/2)} \int_0^\infty \frac{\exp(-xv) }{(1+v)\left(v(2+v)\right)^{1/2-\nu}} dv
\end{equation}
Expanding $1/(1+v)/(2+v)^{1/2-\nu}$ in a Taylor series and integrating yields:
\begin{equation}
\int_x^\infty f_\nu(z) dz =  \frac{\sqrt{\pi}\exp(-x)}{2^{\nu-1/2}\Gamma(\nu+1)\Gamma(\nu+1/2)} \sum \frac{(-1)^j\Gamma\left(\nu + \frac{1}{2} + j\right)}
{(x)^{1/2-\nu-j}} \left[1 + \frac{w_j(1/2-\nu)}{2^j j!}\right]
\end{equation}
where $w_0(\mu) = 0$, $w_1(\mu) = \mu$, $w_2(\mu) = 5\mu + \mu^2$, $w_3(\mu) = 32 \mu + 9 \mu^2 + \mu^3$, and 
$w_4(\mu) = 262 \mu + 83 \mu^2 + 14\mu^3+ \mu^4$.
Using \citet{AS1972} equation 9.6.28,
\begin{equation}
\left(\frac{1}{z}\frac{d}{dz}\right)^m f_{\nu}(z) = \left(\frac{-1}{2}\right)^m \frac{\Gamma(\nu-m+1)}{\Gamma(\nu+1)} f_{\nu-m}(z)
\end{equation}

\end{document}